\documentclass[showpacs,preprintnumbers,amsmath,amssymb,superscriptaddress]{revtex4}
\usepackage{mathrsfs}

\usepackage{amsmath,amssymb,graphicx,bm}

\begin{document}

%\preprint{APS/123-QED}

\title{Discriminating dark energy models by using the statefinder hierarchy and the growth rate of matter perturbations}

\author{Jun Li}
 \affiliation{College of Physical Science and Technology, Hebei University, Baoding 071002, China}

\author{Rongjia Yang\footnote{Corresponding author}}
\email{yangrj08@gmail.com} \affiliation{College of Physical Science
and Technology, Hebei University, Baoding 071002, China}

\author{Bohai Chen}
 \affiliation{College of Physical Science and Technology, Hebei University, Baoding 071002, China}

\date{\today}

\begin{abstract}
We apply the Statefinder hierarchy and the growth rate of matter perturbations to discriminate modified Chaplygin gas (MCG), generalized Chaplygin gas (GCG), superfluid Chaplygin gas (SCG), purely kinetic k-essence (PKK), and $\Lambda$CDM model. We plot the evolutional trajectories of these models in the statefinder plane and in the composite diagnostic plane. We find that GCG, MCG, SCG, PKK, and $\Lambda$CDM can be distinguished well from each other at the present epoch by using the composite diagnostic $\{\epsilon(z), S^{(1)}_{5}\}$. Using other combinations, such as $\{S^{(1)}_{3}, S^{(1)}_4\}$, $\{S^{(1)}_{3}, S_{5}\}$, $\{\epsilon(z), S^{(1)}_{3}\}$, and $\{\epsilon(z), S_4 \}$, some of these five dark energy models cannot be distinguished.
\end{abstract}


\maketitle

\section{Introduction}
Observations of supernovae, large scale structures, and the cosmic microwave background have confirmed that the Universe is experiencing accelerated
expansion. An additional component in the matter sector, dubbed as dark energy, is usually introduced to explain this phenomenon in the framework of general relativity. The simplest and most theoretically appealing candidate of
dark energy is the vacuum energy with a constant equation of state (EoS) parameter $w=-1$ ($\Lambda$CDM). This scenario is consistent
with most of the current astronomical observations but suffers from the cosmological constant problem \cite{weinberg,Carroll2001} and age problem (which is still an open problem) \cite{Yang2010}. It is possible that other unknown matters may be responsible for the accelerated
expansion of the Universe. Over the past years, numerous dark energy models have been proposed, such as quintessence, phantom, k-essence, tachyon, (generalized) Chaplygin gas ((G)CG), etc.

As more and more dark energy models have been proposed, it is important to discriminate various dark energy models. A geometrical diagnostic, called statefinder, is introduced in \cite{V,U}. It has been used to distinguish a number of dark energy models, such as $\Lambda$CDM, quintessence
\cite{V, Sirichai,Linder}, GCG \cite{Gorini,Writambhara,Li}, DGP \cite{Grigoris,Myrzakulov}, Galileon-modified gravity \cite{Sami,Myrzakulov},  purely kinetic k-essence model (PKK) \cite{Gao2010}, holographic dark energy \cite{Granda,Zhang}, Ricci Dark Energy model \cite{Feng}, Agegraphic Dark Energy Model
\cite{Wei}, quintom dark energy model \cite{Wu}, and spatial Ricci scalar dark energy (SRDE) \cite{Yang2012}.

In addition to statefinder, another method, called $Om(z)$ diagnostic \cite{V2}, has been proposed to distinguish dark
energy models. $Om(z)$ is constructed from the Hubble parameter and provides a null test of the $\Lambda$CDM model, namely, if the value of $Om(z)$
is identical at different redshift, then dark energy is $\Lambda$ precisely. $Om(z)$ has been used to compare $\Lambda$CDM with some dark energy models, such as quintessence\cite{V2}, phantom\cite{V2}, PKK \cite{Gao2010}, holographic dark energy \cite{Granda}, and SRDE \cite{Yang2012}. A simple extension of the $Om(z)$ diagnostic, called $Om3(z)$, combining standard ruler information from BAO with standard candle information from type Ia supernovae, yield a powerful novel null diagnostic of the cosmological constant hypothesis \cite{Shafieloo2012}. It has been shown, however, that one cannot distinguish PKK from the $\Lambda$CDM model at $68.3\%$ confidence level by using $Om(z)$ and the statefinder $\{r, s\}$ \cite{Gao2010}. So it is natural to look for new methods to distinguish PKK (or other dark energy models) from the $\Lambda$CDM model.

In \cite{Maryam}, it has been shown that the Statefinder hierarchy combined with the growth rate of matter perturbations defines a composite null diagnostic which can distinguish DGP from the $\Lambda$CDM model. So it is natural to ask whether it can distinguish PKK from the $\Lambda$CDM model?

In this paper, we use the statefinder hierarchy and the growth rate of matter perturbations proposed in \cite{Maryam} to distinguish GCG, modified Chaplygin gas (MCG), superfluid Chaplygin gas (SCG), and PKK from $\Lambda$CDM model. We find that GCG, MCG, SCG, and PKK can be discriminated from the $\Lambda$CDM model by using the statefinder hierarchy and the growth rate of matter perturbations.

The rest of the paper is organized as follows. In Sec. II, we will briefly review the Statefinder hierarchy and the growth rate of matter perturbations. In Sec. III, we use the Statefinder hierarchy and the growth rate of matter perturbations to distinguish GCG, MCG, SCG, and PKK from the $\Lambda$CDM model. In the last section some conclusions and discussions are presented.

\section{The Statefinder hierarchy and the growth rate of matter perturbations}
In this section, we first introduce the usual formalism of Statefinder hierarchy, and then briefly describe the growth rate of matter perturbations.
\subsection{The Statefinder hierarchy}
Like the Statefinder $\{r, s\}$ which is related to the third derivative of the expansion factor and is defined as \cite{V,U}
\begin{eqnarray}
r\equiv \frac{\dddot{a}}{aH^3},~~~~s\equiv\frac{r-1}{3(q-1/2)},
\end{eqnarray}
where $H=\dot{a}/a$ is the Hubble parameter and $q\equiv -\ddot{a}/(aH^2)$ is the deceleration parameter, the ``Statefinder hierarchy" includes higher derivatives of the scale factor $d^{n}a/dt^{n}$, $n\geq2$. It has been demonstrate that all members of the Statefinder hierarchy can be expressed in terms of elementary functions of the deceleration parameter $q$ (equivalently the matter energy density parameter $\Omega_{\rm m}(z)=8\pi G\rho_{\rm m}/(3H^2)$) \cite{Maryam}.

The scale factor $a(t)/a_{0}=(1+z)^{-1}$ can be Taylor expanded around the present time $t_{0}$ as follows:
\begin{eqnarray}
\frac{a(t)}{a_{0}}=1+\sum^{\infty}_{n=1}\frac{A_{n}(t_{0})}{n!}[H_{0}(t-t_{0})]^{n},
\end{eqnarray}
where
\begin{eqnarray}
A_{n}\equiv\frac{a^{(n)}}{aH^{n}},
\end{eqnarray}
with $a^{(n)}=d^na/dt^n$ and $n$ is a positive integer. Historically $-A_2$ has been called deceleration parameter $q$, $A_3$ is the Statefinder $r$ \cite{V,U} or the jerk $j$ \cite{Visser}, $A_4$ is the snap \cite{Visser,Capozziello,Dabrowski,Dunajski} and $A_5$ the lerk $l$ \cite{Visser,Capozziello,Dabrowski,Dunajski}. For $\Lambda$CDM in a spatially flat, homogeneous, and isotropic Friedmann-Lema\^{i}tre-Robertson-Walker (FLRW) universe, we can easily get:
\begin{align}
&A_{2}=1-\frac{3}{2}\Omega_{\rm m},\\
&A_{3}=1,\\
&A_{4}=1-\frac{3^2}{2}\Omega_{\rm m},\\
&A_{5}=1+3\Omega_{\rm m}+\frac{3^3}{2}\Omega_{\rm m}^{2},....
\end{align}
For $\Lambda$CDM we have $\Omega_{m}=\frac{2}{3}(1+q)$. The Statefinder hierarchy, $S_{\rm n}$, is defined as \cite{Maryam}:
\begin{align}
&S_{2}=A_{2}+\frac{3}{2}\Omega_{\rm m},\\
&S_{3}=A_{3},\\
&S_{4}=A_{4}+\frac{3^2}{2}\Omega_{\rm m},\\
&S_{5}=A_{5}-3\Omega_{\rm m}-\frac{3^3}{2}\Omega_{\rm m}^{2},....
\end{align}
The Statefinder hierarchy is unchanged for $\Lambda$CDM during the expansion of the universe since \cite{Maryam}
\begin{eqnarray}
\label{Sn}
S_{n}|_{\Lambda \rm{CDM}}=1,~~~~{\rm{for}}~~~~ n=2,3,4,....
\end{eqnarray}
These equations define a null diagnostic for $\Lambda$CDM, since for evolving dark energy models some of these equalities in (\ref{Sn}) may be violated. When $n\geq3$, more than one way can be adopted to define a null diagnostic, see for example \cite{Maryam} one series of Statefinders can be defined as:
\begin{align}
&S^{(1)}_{3}=S_{3},\\
&S^{(1)}_{4}=A_{4}+3(1+q)\\
&S^{(1)}_{5}=A_{5}-2(4+3q)(1+q), ....
\end{align}
This series of Statefinders also stays at unity for $\Lambda$CDM during the expansion of the universe \cite{Maryam}
\begin{eqnarray}
S^{(1)}_{n}|_{\Lambda \rm{CDM}}=1.
\end{eqnarray}
In other words, $\{S_{n}, S^{(1)}_{n}\}|_{\Lambda \rm{CDM}}=\{1,1\}$. For other dark energy models, however, the pair $\{S_{n}, S^{(1)}_{n}\}$ are expected to give different results.

A second member of the Statefinder hierarchy could be constructed from the Statefinder $S^{(1)}_{n}$ as
follows \cite{Maryam}:
\begin{eqnarray}
\label{sn}
S^{(2)}_{n}=\frac{S^{(1)}_{n}-1}{3(q-1/2)}.
\end{eqnarray}
For $\Lambda$CDM, $S^{(1)}_{n}=1$ and $S^{(2)}_{n}=0$, namely $\{S_{n}, S^{(2)}_{n}\}|_{\Lambda \rm{CDM}}=\{1,0\}$. The pair $\{S_{n}, S^{(2)}_{n}\}$ are also expected to give different results for other dark energy models.

Consequently the Statefinder hierarchy $\{S_{n}, S^{(1)}_{n}\}$ or $\{S_{n}, S^{(2)}_{n}\}$ can be used as an excellent means of distinguish dynamical dark energy models from $\Lambda$CDM. In \cite{Maryam}, CG, DPG, $w$CDM (models with a constant EoS), and $\Lambda$CDM have been distinguished by using $\{S_3^{(1)}, S_4^{(1)}\}$.
%%%%%%%%%%%%%%%%%%%
\begin{figure}
\includegraphics[width=9cm]{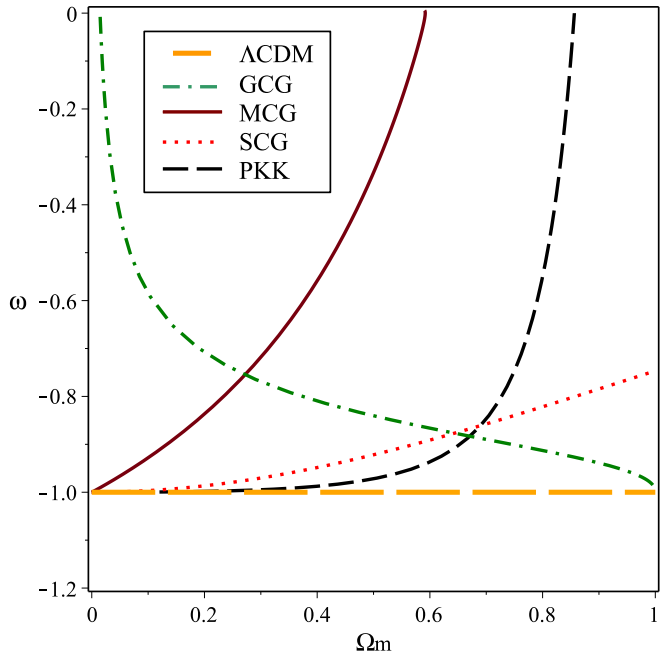}
\caption{Evolution trajectories in the $\Omega_{\rm m}-w$ plane for five dark energy models. The long-dash, solid, dash, dot, dash-dot line represents the evolution of $\Lambda$CDM, MCG, PKK, SCG, and GCG, respectively. \label{omw}}
\end{figure}
%%%%%%%%%%%%%%%%%

\subsection{The growth rate of matter perturbations}
In this subsection, we will examine the linearized density perturbation. The equation of the linear matter density contrast, $\delta=\delta\rho_{\rm m}/\rho_{\rm m}$, is given by
\begin{eqnarray}
\label{pert}
\ddot{\delta}+2H\dot{\delta}=4\pi G\rho_{\rm m}\delta,
\end{eqnarray}
where the dot denotes a derivative with respect to time $t$. This equation holds only if dark energy
arises because of modifications to the right hand side of the Einstein equation and in the absence of anisotropic stress. In this case Eq. (\ref{pert}) holds on scales much smaller than the Jeans length for dark energy perturbations. The matter perturbation $\delta$ contains essentially the same information as the expansion rate $H(z)$ \cite{LP,Alam2009}.

The evolution equations of background in a flat universe are:
\begin{eqnarray}
\label{frei}
H^2=\frac{8\pi G}{3}(\rho_{\rm m}+\rho_{\rm DE}),
\end{eqnarray}
\begin{eqnarray}
2\frac{\ddot{a}}{a}+H^2=-8\pi G w\rho_{\rm DE},
\end{eqnarray}
where $w$ is the EoS of dark energy: $w\equiv p_{\rm DE}/\rho_{\rm DE}$. Using the definition of matter energy density parameter, we can derive
\begin{eqnarray}
\label{acc}
\frac{\ddot{a}}{a}=-\frac{1}{2}H-\frac{3}{2}w(1-\Omega_{\rm m})H.
\end{eqnarray}

By combining Eqs. (\ref{pert}) and (\ref{acc}) we rewrite the linear perturbation equation as \cite{LP}
\begin{eqnarray}
\label{per1}
\frac{d^2\ln\delta}{d\ln a^2}+\left(\frac{d\ln\delta}{d\ln a}\right)^2+\left[\frac{1}{2}-\frac{3}{2}w(1-\Omega_{\rm m})\right]\frac{d\ln\delta}{d\ln a}=\frac{3}{2}\Omega_{\rm m}.
\end{eqnarray}

The fractional growth parameter $\epsilon(z)$ used with statefinder can be defined as follows \cite{VADS}:
\begin{eqnarray}
\label{12}\epsilon(z)=\frac{f(z)}{f_{\Lambda \rm{CDM}}(z)},
\end{eqnarray}
where$f(z)=d \ln\delta/d\ln a$ describes the growth rate of linear density perturbations \cite{LP} and can be parameterized as
\begin{eqnarray}
\label{13}f(z)\simeq\Omega^{\gamma}_{\rm m}(z),
\end{eqnarray}
where $\gamma(z)$ is the growth index parameter. The expression of $\gamma(z)$ can be obtained as follows.

Using Eqs. (\ref{frei}) and conservation of the stress energy, we have
\begin{eqnarray}
\label{om}
\frac{d\Omega_{\rm m}}{d\ln a}=3w(1-\Omega_{\rm m})\Omega_{\rm m}.
\end{eqnarray}
Combining Eqs. (\ref{per1}) and (\ref{om}), we obtain the equation for $f$ in terms of $\Omega_{\rm m}$ \cite{LP}
\begin{eqnarray}
\label{per2}
3w\Omega_{\rm m}(1-\Omega_{\rm m})\frac{df}{d\Omega_{\rm m}}+\left[\frac{1}{2}-\frac{3}{2}w(1-\Omega_{\rm m})\right]f+f^2=\frac{3}{2}\Omega_{\rm m}.
\end{eqnarray}
Taking into account Eq. (\ref{13}), Eq. (\ref{per2}) becomes \cite{LP}
\begin{eqnarray}
\begin{split}
3w(1-\Omega_{\rm m})\Omega_{\rm m}\ln\Omega_{\rm m}\frac{d\gamma}{d\Omega_{\rm m}}-3w(\gamma-\frac{1}{2})\Omega_{\rm m}+\Omega_{\rm m}^{\gamma}
-\frac{3}{2}\Omega_{\rm m}^{1-\gamma}+3w\gamma-\frac{3}{2}w
+\frac{1}{2}=0.
\end{split}
\end{eqnarray}
For slowly varying EoS which satisfies $|dw/d\Omega_{\rm m}|\ll (1-\Omega_{\rm m})^{-1}$, we can have \cite{LP}
\begin{eqnarray}
\label{14}
\gamma(z)\simeq\frac{3}{5-\frac{w}{1-w}}+\frac{3}{125}\frac{(1-w)(1-\frac{3}{2}w)}{(1-\frac{6}{5}w)^{3}}(1-\Omega_{\rm m}).
\end{eqnarray}
The above approximation works reasonably well for physical dark energy models with a constant or a slowly variational EoS. For $\Lambda$CDM, it is obvious that
\begin{eqnarray}
\epsilon(z)|_{\Lambda \rm{CDM}}=1.
\end{eqnarray}
So we can combine the fractional growth parameter $\epsilon(z)$ and the Statefinders to define a composite null diagnostic: $\{\epsilon(z), S_{n}\}$ or $\{\epsilon(z), S^{(m)}_{n}\}$ \cite{Maryam}. Using $\{\epsilon(z), S_3^{(1)}\}$, DPG, $w$CDM, and $\Lambda$CDM have been distinguished in \cite{Maryam}.
%%%%%%%%%%%%%%%%%%%
\begin{figure}
\includegraphics[width=9cm]{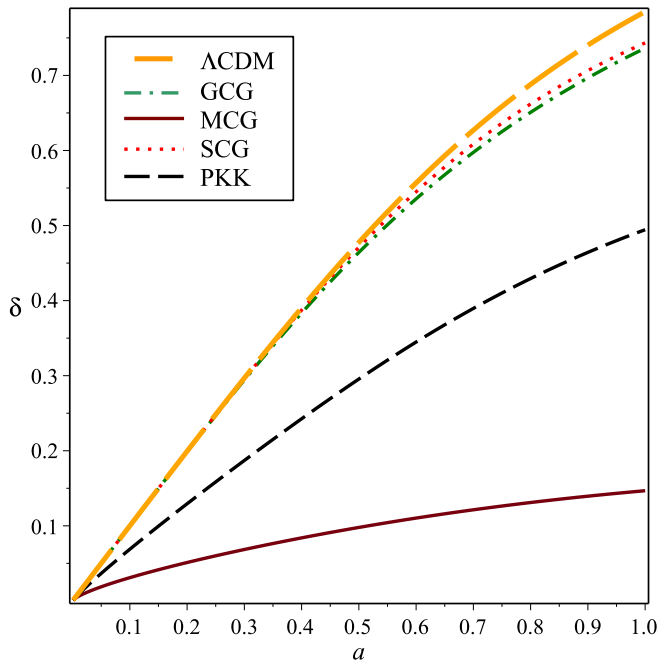}
\caption{The evolutions of the linear matter density contrast $\delta$ for $\Lambda$CDM, MCG, PKK, SCG, and GCG models. \label{per}}
\end{figure}
%%%%%%%%%%%%%%%%%%%

\section{Dark energy models and discriminations}
In this section, we will use the Statefinder hierarchy and the growth rate of matter perturbations to distinguish some dark energy models.
$\Lambda$CDM, $w$CDM, CG, and DGP models have been distinguished by using these two methods in Ref. \cite{Maryam}. Here our discussions focus on the following models: $\Lambda$CDM, generalized Chaplygin gas (GCG), modified Chaplygin gas (MCG), superfluid Chaplygin gas (SCG), and purely kinetic k-essence model (PKK). For simplicity we neglect the contributions of radiation and spatial curvature.

\subsection{Dark energy models}
In Chaplygin gas unifying dark matter and dark energy, such as CG, GCG, and MCG, the negligible sound speed may produce unphysical oscillations and an exponential blowup in the dark matter power spectrum \cite{Sandvik}. This problem can be solved by decomposing the energy density into dark matter and dark energy \cite{Bento}. So here we treat GCG and MCG only as dark energy, not as models unifying dark matter and dark energy.

(1) GCG as dark energy is characterized with $p=-A/\rho^\alpha$ with $A$ a positive constant and $0<\alpha\leq 1$. For $\alpha=1$, GCG reduces to the CG model. The action for GCG can be written as a generalized Born-Infeld form \cite{Bento2002,Gorini2004,Chimento} whose parameters has been constrained with observational data in \cite{Yang2008}. The EoS of GCG and the expansion rate of a universe containing the GCG dark energy and pressureless matter are given by, respectively
\begin{eqnarray}
\label{5}w=-\frac{A_{\rm s}}{A_{\rm s}+(1-A_{\rm s})(1+z)^{3(1+\alpha)}},
\end{eqnarray}
\begin{eqnarray}
\label{4}E(z)=\frac{H(z)}{H_0}=\bigg[\Omega_{\rm m0}(1+z)^3+(1-\Omega_{\rm m0}) \bigg(A_{\rm s}+(1-A_{\rm s})(1+z)^{3(1+\alpha)}\bigg)^{\frac{1}{1+\alpha}}\bigg]^{\frac{1}{2}}.
\end{eqnarray}
where $A_{\rm s}=A/\rho^{1+\alpha}_{\rm{gcg0}}$ with $\rho_{\rm{gcg0}}$ being the present value of the energy density of GCG. GCG as dark energy has been successfully confronted with observational tests \cite{GSA}. The values of parameters we take in the following content are constrained from large-scale structure observation: $A_{\rm s}=0.764$, $\alpha=-1.436$, and $\Omega_{\rm m0}=0.2895$ \cite{GSA}.

(2) MCG is considered as dark energy with $P=B\rho-A/\rho^\alpha$, where $A$, $B$, and $\alpha$ are positive constants with $0<\alpha\leq 1$. For $B=0$, MCG reduces to the GCG model. For $B=0$ and $\alpha=1$, MCG reduces to the CG model. For $A=0$, MCG reduces to dark energy models with a constant EoS $w=B$. The EoS of MCG and the normalized Hubble parameter take the form, respectively
\begin{eqnarray}
\label{7}w=B-\frac{A_{\rm s}(1+B)}{A_{\rm s}+(1-A_{\rm s})(1+z)^{3(1+B)(1+\alpha)}},
\end{eqnarray}
\begin{eqnarray}
\label{6}E(z)=\bigg[\Omega_{\rm m0}(1+z)^3+(1-\Omega_{\rm m0})\bigg(A_{\rm s}+(1-A_{\rm s})(1+z)^{3(1+B)(1+\alpha)}\bigg)^{\frac{1}{1+\alpha}}\bigg]^{\frac{1}{2}}.
\end{eqnarray}
where $A_{\rm s}=A/(1+B)\rho^{1+\alpha}_{\rm{mcg0}}$ with $\rho_{\rm{mcg0}}$ being the present value of the energy density of MCG. Compared with GCG the proposed MCG is suitable to describe the evolution of the universe over a wide range of epoch \cite{UAS}. The best-fit values of parameters we take are: $A_{\rm s}=0.769$, $B=0.008$, $\alpha=0.002$, and $\Omega_{\rm m0}=0.262$ \cite{BP}.

(3) SCG is a model unifying dark matter and dark energy \cite{VAP}. It involves a Bose-Einstein condensate as dark energy possessing the EoS of Chaplygin gas and an excited state acts as dark matter. Though the component of dark energy possesses the
EoS of the Chaplygin gas, but the evolution
of the universe provided by SCG is different from
that in the two-component model with the Chaplygin gas
and cold dark matter as well as from that in the GCG unifying dark energy and dark matter \cite{VAP, Yang2013}.
The EoS of SCG and the the normalized Hubble parameter are given by, respectively
\begin{eqnarray}
\label{9}w=-\frac{\big[(1+z)^{-3}+k_{0}\big]^2}{k^{2}+\big[(1+z)^{-3}+k_{0}\big]^{2}},
\end{eqnarray}
\begin{eqnarray}
\label{8}E(z)=(1+z)^3\bigg[\frac{k^2+\big(k_0+(1+z)^{-3}\big)^2}{k^2+(1+k_0)^2}\bigg]^{\frac{1}{2}}.
\end{eqnarray}
where $k$ and $k_0$ are positive constants. $k$ gives an initial normalized total particle number density. $k_0$ is associated to the ratio of normal and condensate density evaluated at present. The best-fit values of parameters we take are: $k=0.173$ and $k_{0}=0.297$ \cite{RAV}.

(4) PKK is a class of k-essence with Lagrangian: $p(X)=-V_0\sqrt{1-2X}$, where $X\equiv \frac{1}{2}\partial_{\mu}\phi\partial^{\mu}\phi$ is the the kinetic energy
and $V_0$ is a constant potential \cite{Chim04,yang08}. The expansion rate of a universe containing the PKK dark energy and pressureless matter takes the form \cite{yang09}
\begin{eqnarray}
\label{10}E(z)=\bigg[\Omega_{\rm m0}(1+z)^3+(1-\Omega_{\rm m0})f(z)\bigg]^{\frac{1}{2}},
\end{eqnarray}
where
\begin{eqnarray}
f(z)=\exp\bigg[3\int_{0}^{z}\frac{1+w(z')}{1+z'}dz'\bigg],
\end{eqnarray}
with EoS
\begin{eqnarray}
\label{11}
w=-\frac{1}{1+2k_{0}^{2}(1+z)^{6}},
\end{eqnarray}
where $k_0$ is a constant. For $k_0=0$, the above EoS
reduces to $-1$; meaning the $\Lambda$CDM model is contained in
PKK model as one special case. The behavior of the EoS
(\ref{11}), runs closely to zero in the early Universe, while runs closely to
$-1$ in the future. The best-fit values of parameters we take are: $\Omega_{\rm m0}=0.36$ and $k_{0}=0.067$ \cite{yang09}.
%%%%%%%%%%%%%%%
\begin{figure}
\includegraphics[width=9cm]{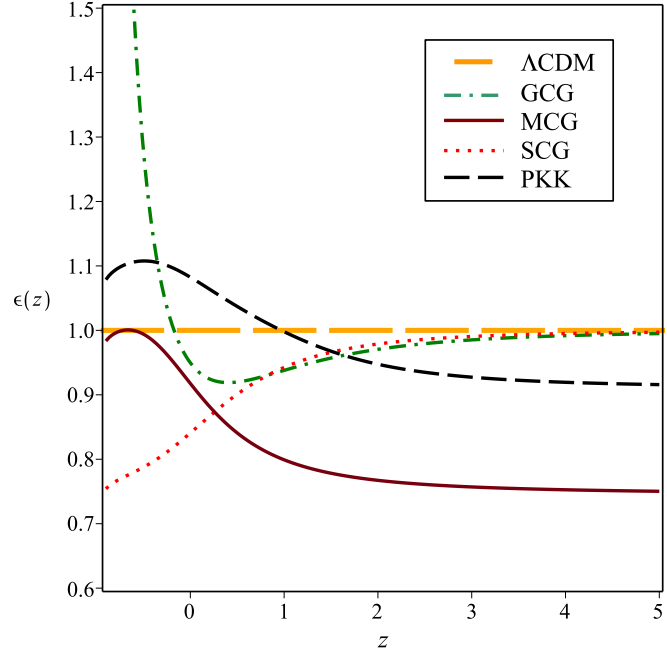}
\caption{The evolutions of the fractional growth parameter $\epsilon(z)$ for $\Lambda$CDM, MCG, PKK, SCG, and GCG models. \label{ez}}
\end{figure}
%%%%%%%%%%%%%%%%%

\subsection{Discriminations with the Statefinder hierarchy and the growth rate of matter perturbations}
Now we will use Statefinder hierarchy and the growth rate of matter perturbations to distinguish dark energy models described above. The deceleration parameter $q$, $A_3$, $A_4$, and $A_5$ can be rewritten as
\begin{eqnarray}
\label{q}
q &&=(1+z)\frac{1}{E}\frac{dE}{dz}-1,\\
\label{A3}
A_3 &&=(1+z)\frac{1}{E^2}\frac{d[E^2(1+q)]}{dz}-3q-2,\\
\label{A3}
A_4 &&=\frac{-(1+z)}{E^3}\frac{d[E^3(2+3q+A_3)]}{dz}+4A_3+3q(q+4)+6,\\
\label{A5}
A_5 &&=\frac{-(1+z)}{E^4}\frac{d[E^4(A_4-4A_3-3q(q+4)-6)]}{dz}+5A_4-10A_3(q+2)-30q(q+2)-24.
\end{eqnarray}
In order to distinguish dark energy models by using Statefinder hierarchy and the growth rate of matter perturbations, we need another important parameter, the dimensionless matter density parameter. The dimensionless matter density parameter for GCG, MCG, and PKK is $\Omega_{\rm m}=\Omega_{\rm m0}(1+z)^3/E^2$ and for SCG is $\Omega_{\rm m}=k_0/[k_0+(1+z)^{-3}]$.
%%%%%%%%%%%%%%%
\begin{figure}
\includegraphics[width=9cm]{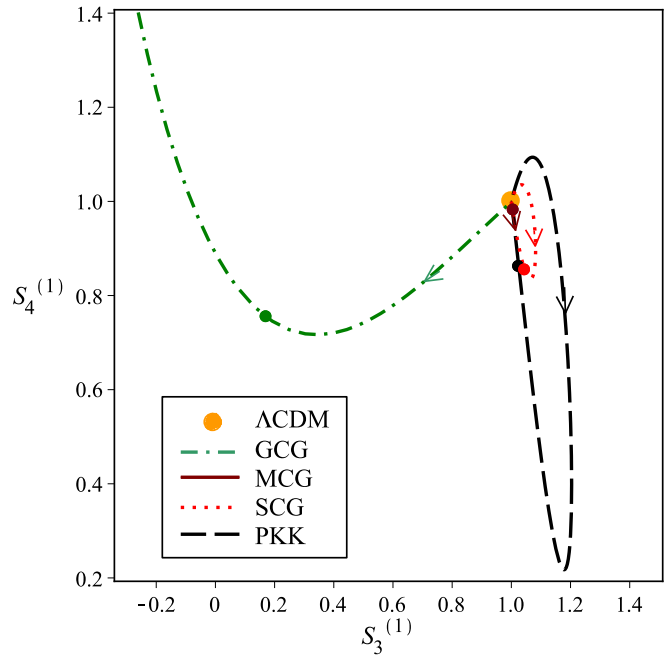}
\caption{The Statefinder $\{S^{(1)}_{3}, S^{(1)}_{4}\}$ are shown for $\Lambda$CDM, MCG, PKK, SCG, and GCG models. The
arrows show time evolution and the present epoch in different models is shown as a dot. \label{nS41S31}}
\end{figure}
%%%%%%%%%%%%%%%%%

Before using the fractional growth parameter $\epsilon(z)$ to distinguish dark energy models, we investigate whether the EoSs of MCG, PKK, SCG, and GCG vary slowly so that Eq. (\ref{14}) can be used. In Fig. \ref{omw}, we plot the evolutional trajectories in $\Omega_{\rm m}-w$ plane for five dark energy models considered here. The EoS of PKK varies very slowly at the present epoch but varies fast when $\Omega_{\rm m}\geq 0.7$. The EoS of SCG varies more quickly than that of PKK at the present epoch but varies more slowly than that of PKK when $\Omega_{\rm m}\geq 0.7$. The EoS of GCG varies quickly when $\Omega_{\rm m}\leq 0.2$ and then varies more slowly when $\Omega_{\rm m}$ increases. The EoS of MCG varies more quickly than those of PKK, SCG, and GCG when $\Omega_{\rm m}$ increases from $0.2$ to $0.4$. For $z=0$, we find $|dw/d\Omega_{\rm m}|=0.079\ll 1/(1-\Omega_{\rm m})\simeq 1.56$ for PKK, $|dw/d\Omega_{\rm m}|=0.152< 1/(1-\Omega_{\rm m})\simeq 1.297$ for SCG, $|dw/d\Omega_{\rm m}|=0.495< 1/(1-\Omega_{\rm m})\simeq 1.41$ for GCG, and $|dw/d\Omega_{\rm m}|=1.22< 1/(1-\Omega_{\rm m})\simeq 1.36$ for MCG. Obviously the EoSs of MCG and GCG donot vary so slowly that the condition $|dw/d\Omega_{\rm m}|\ll 1/(1-\Omega_{\rm m})$ is not satisfied well by them, though the best-fit values of parameters of MCG and GCG we take are constrained from observational growth data by using the Eq. (\ref{14}) (under the condition $|dw/d\Omega_{\rm m}|\ll 1/(1-\Omega_{\rm m})$)\cite{GSA,BP}.
%%%%%%%%%%%%%%%%%%%%
\begin{figure}
\includegraphics[width=9cm]{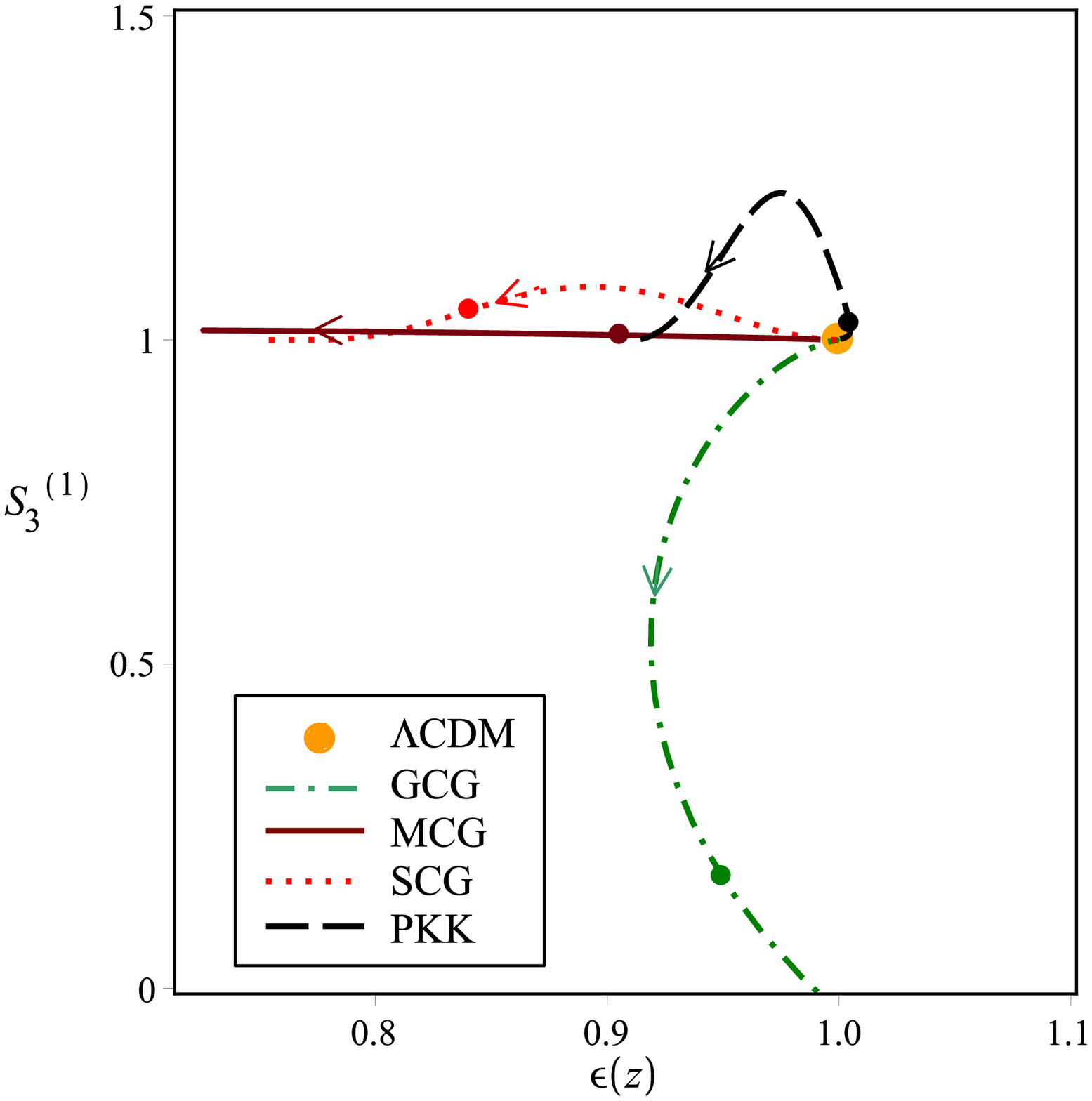}
\caption{The composite diagnostic $\{\epsilon(z), S_{3}^{(1)}\}$ is plotted for $\Lambda$CDM, MCG, PKK, SCG, and GCG models. The
arrows show time evolution and the present epoch in different models is shown as a dot. \label{S31e}}
\end{figure}
%%%%%%%%%%%%%%%%%%%%%%

In Fig. \ref{per}, we solve Eq. (\ref{per1}) numerically for a relevant choice of parameters. The deviations of the linear matter density contrast among $\Lambda$CDM, SCG, and GCG are very small, even smaller than $0.1$. But the deviations of the linear matter density contrast among $\Lambda$CDM, PKK, and MCG are large, especially the growth of the linear matter density contrast $\delta$ for MCG increases very slowly.

In Fig. \ref{ez}, we plot the evolutions of the fractional growth parameter $\epsilon(z)$ for $\Lambda$CDM, MCG, PKK, SCG, and GCG models. Obviously the evolution of $\epsilon(z)$ for MCG is significantly different from that of $\Lambda$CDM, but those of SCG and GCG run closely to that of $\Lambda$CDM for $z>2$, meaning that SCG, GCG and $\Lambda$CDM cannot be distinguished from each other at hight redshift. At the present epoch ($z=0$), the deviations of $\epsilon(z)$ among $\Lambda$CDM, MCG, PKK, SCG, and GCG can be discriminated. These results for $\epsilon(z)$ match the results obtained by solving Eq. (\ref{per1}) numerically.
%%%%%%%%%%%%%%%%%%
\begin{figure}
\includegraphics[width=9cm]{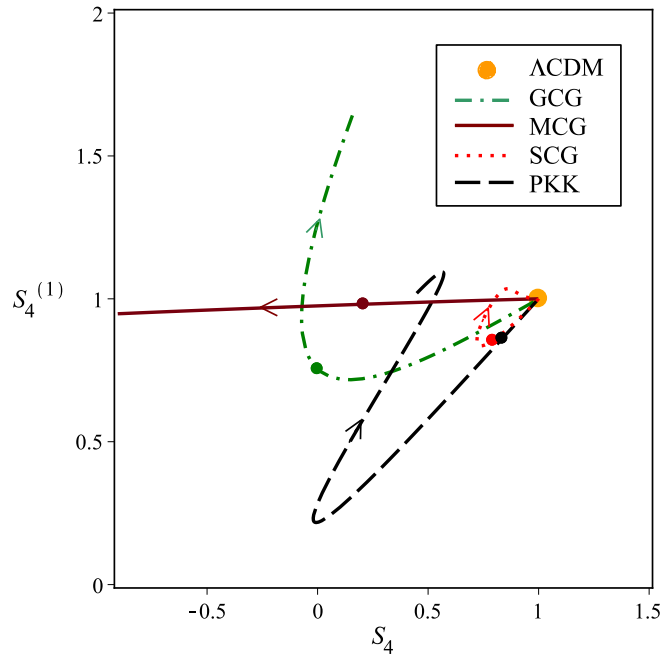}
\caption{The Statefinder $\{S_4, S^{(1)}_{4}\}$ are shown for $\Lambda$CDM, MCG, PKK, SCG, and GCG models. \label{kS41S4}}
\end{figure}
%%%%%%%%%%%%%

In \cite{Maryam}, it has been shown that CG, DPG, $w$CDM, and $\Lambda$CDM can be distinguished by using $\{S_3^{(1)}, S_4^{(1)}\}$ and DPG, $w$CDM, and $\Lambda$CDM can be distinguished by using $\{\epsilon(z), S_3^{(1)}\}$. We also use the Statefinder $\{S_3^{(1)}, S_4^{(1)}\}$ to discriminate GCG, MCG, SCG, PKK, and $\Lambda$CDM. We plot the evolutional trajectories of these models in the $\{S_3^{(1)}, S_4^{(1)}\}$ plane. It is obvious that the trajectory of GCG is very different from those of MCG, SCG, PKK, and $\Lambda$CDM, and it can be distinguished well from them at the present epoch. SCG and PKK may be distinguished from $\Lambda$CDM at a certain epoch but they cannot be distinguished from each other at the present epoch. MCG and $\Lambda$CDM also cannot be distinguished from each other at the present epoch by using $\{S_3^{(1)}, S_4^{(1)}\}$, as shown in Fig. \ref{nS41S31}.
%%%%%%%%%%%%
\begin{figure}
\includegraphics[width=9cm]{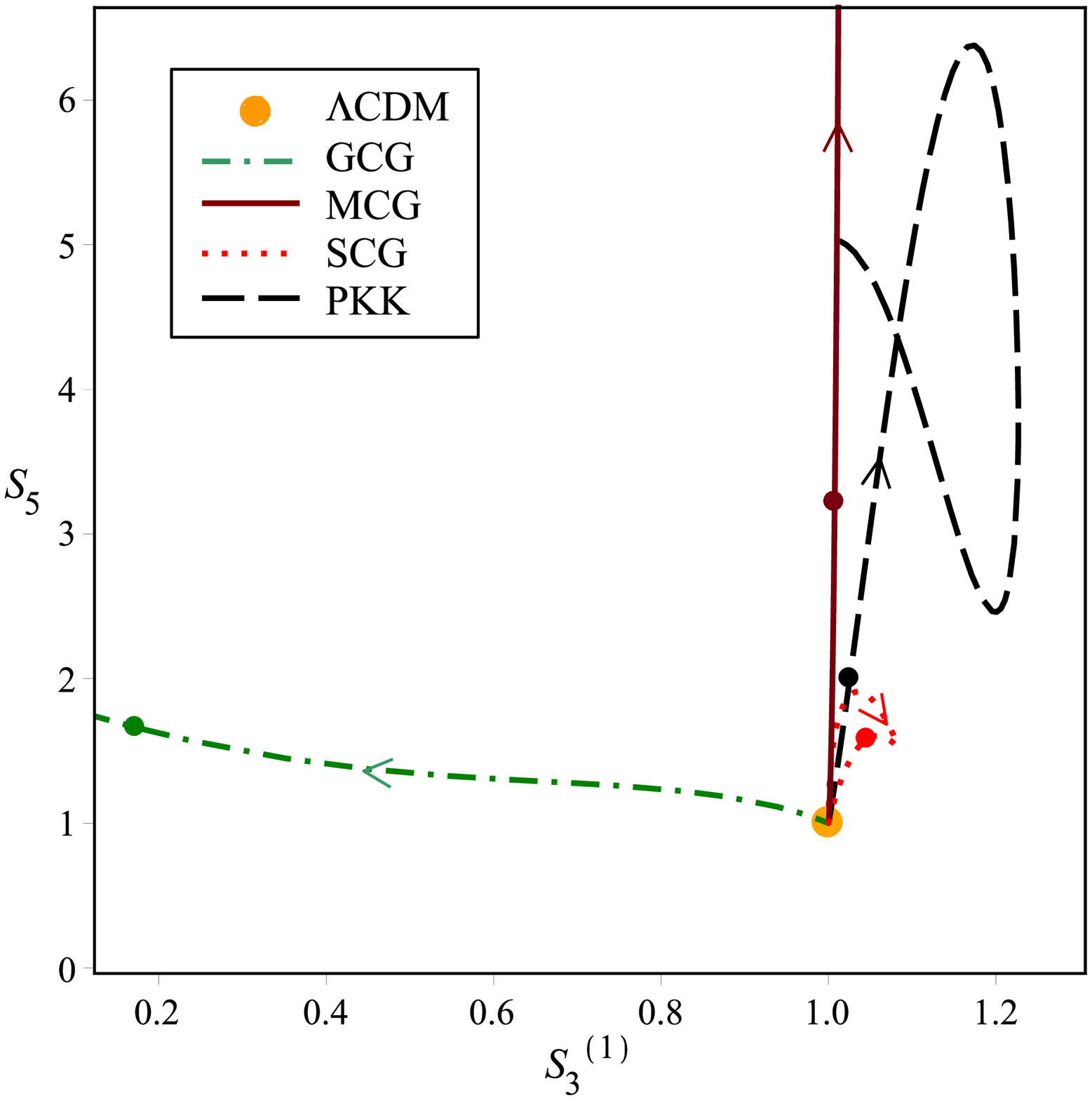}
\caption{The Statefinder $\{S_{3}^{(1)}, S_5\}$ are shown for $\Lambda$CDM, MCG, PKK, SCG, and GCG models. \label{S31S5}}
\end{figure}
%%%%%%%%%%%%%

Using the composite diagnostic $\{\epsilon(z), S_3^{(1)}\}$ to distinguish these dark energy models, we find that at the present epoch GCG, MCG, SCG can be distinguished well from $\Lambda$CDM or PKK, but $\Lambda$CDM and PKK cannot be distinguished from each other, as shown in Fig. \ref{S31e}.

In the Statefinder $\{S_4, S^{(1)}_{4}\}$ plane, it is obvious that the evolutional trajectories of these models are different from each other. We find that at the present epoch MCG and GCG can be distinguished from $\Lambda$CDM, PKK, or SCG; but PKK and SCG cannot be distinguished from each other, as shown in Fig. \ref{kS41S4}.

Using Statefinder $\{S_{3}^{(1)}, S_5\}$, MCG and GCG can be distinguished from $\Lambda$CDM, PKK, or SCG;  PKK and SCG may can be distinguished from $\Lambda$CDM but cannot be distinguished from each other at present epoch, as shown in Fig. \ref{S31S5}.
%%%%%%%%%%%%%%%%%%%%
\begin{figure}
\includegraphics[width=9cm]{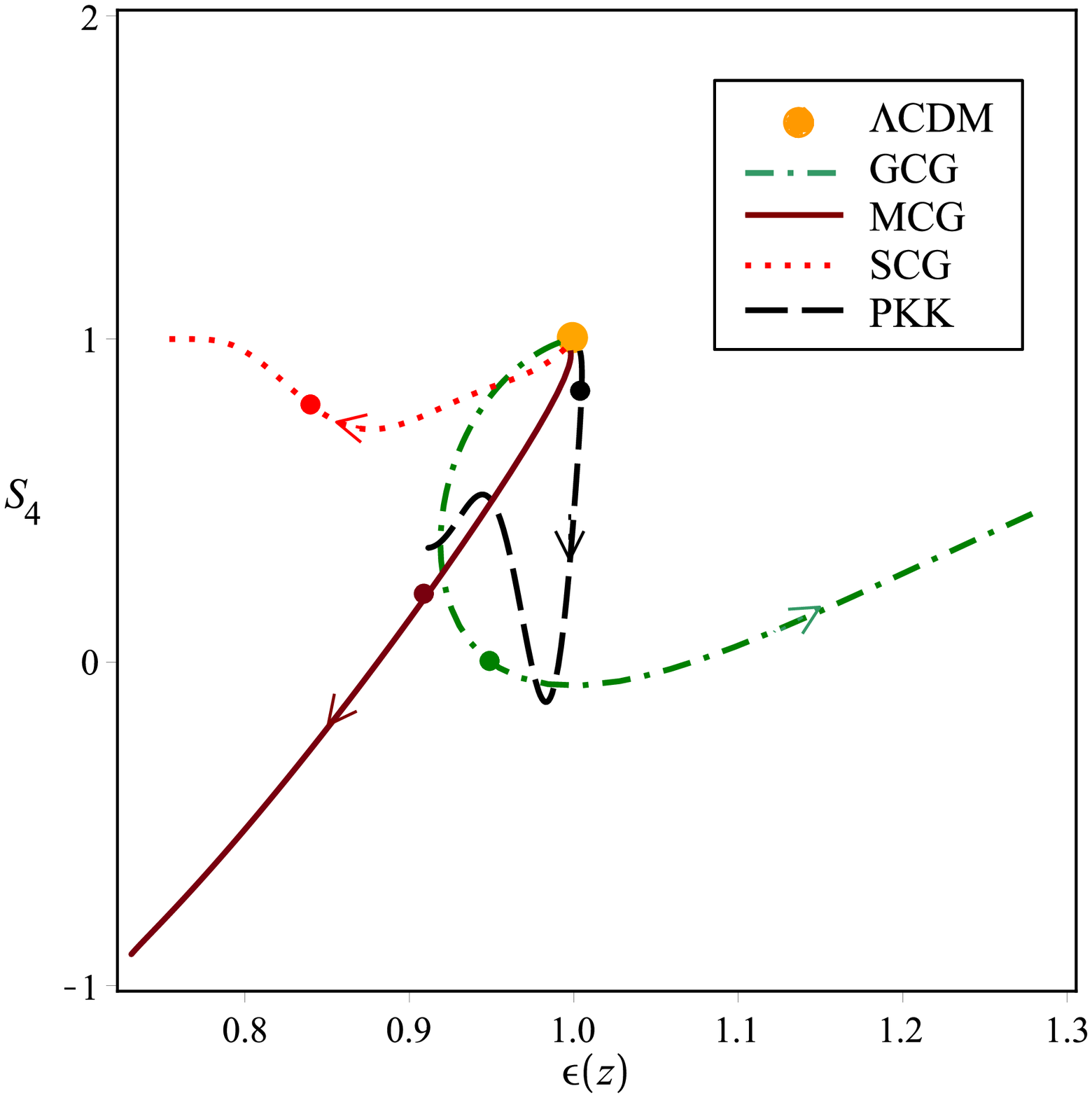}
\caption{The composite diagnostic $\{\epsilon(z), S_4\}$ is plotted for $\Lambda$CDM, MCG, PKK, SCG, and GCG models. \label{S4e}}
\end{figure}
%%%%%%%%%%%%%%

In the composite diagnostic $\{\epsilon(z), S_4\}$ plane, we find that evolutional trajectories of these models are very different from each other; MCG, SCG, and GCG can be distinguished well from PKK or $\Lambda$CDM, but PKK and $\Lambda$CDM cannot be distinguished well from each other at the present epoch, as shown in Fig. \ref{S4e}.

\begin{figure}
\includegraphics[width=9cm]{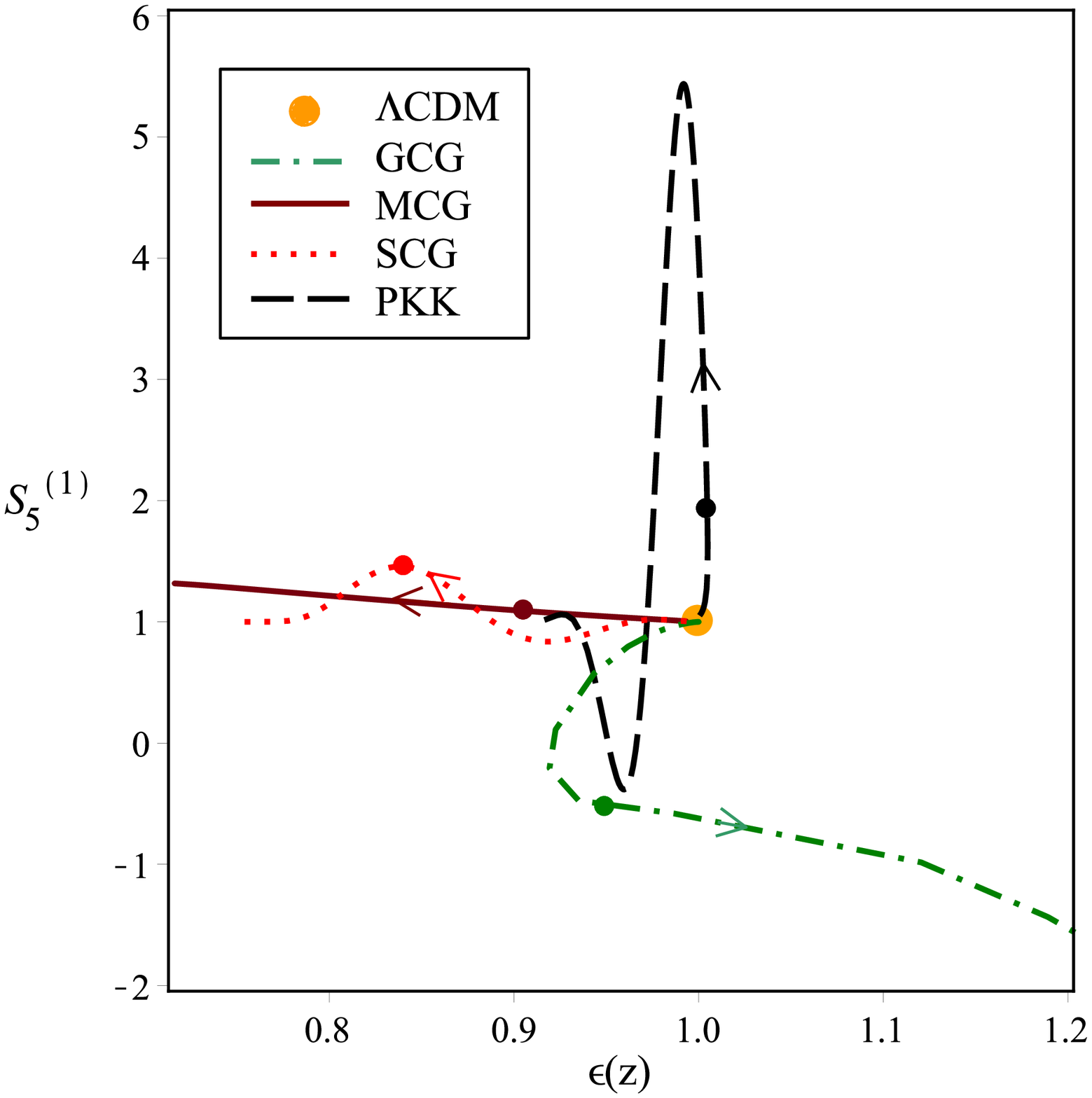}
\caption{The composite diagnostic $\{\epsilon(z), S_{5}^{(1)}\}$ is plotted for $\Lambda$CDM, MCG, PKK, SCG, and GCG models. \label{S51}}
\end{figure}

In Fig. \ref{S51}, we find the trajectories of MCG and SCG are similar but other models' trajectories are different in the composite diagnostic $\{\epsilon(z), S_5^{(1)}\}$ plane, MCG, SCG, GCG, PKK, and $\Lambda$CDM can be distinguished well from each other at the present epoch.
%%%%%%%%%%%%%%%%%%%%%%%%%
%%%%%%%%%%%%%%%%%%

We also use other pairs, such as $\{S_3^{(1)}, S_5^{(1)}\}$, $\{S_4, S_5^{(1)}\}$, $\{S_5, S_4^{(1)}\}$, $\{\epsilon(z), S_4^{(1)}\}$, etc., to distinguish these models and find they cannot be distinguished well, compared with the results obtained by using pairs presented above.

\section{Conclusions and discussions}
We have used the Statefinder hierarchy and the growth rate of matter perturbations to discriminate MCG, GCG, SCG, PKK, and $\Lambda$CDM dark energy models. The evolutional trajectories of these dark energy models in the statefinder hierarchy, such as $\{S^{(1)}_{3}, S^{(1)}_4\}$, $\{S_4, S^{(1)}_{4}\}$, and $\{S^{(1)}_{3}, S_{5}\}$, and in the composite diagnostic, such as $\{\epsilon(z), S^{(1)}_{3}\}$, $\{\epsilon(z), S_4 \}$, and $\{\epsilon(z), S^{(1)}_{5}\}$, have been plotted.

We have found that GCG, MCG, SCG, PKK, and $\Lambda$CDM can only be distinguished well from each other at the present epoch by using the composite diagnostic $\{\epsilon(z), S^{(1)}_{5}\}$. Using the Statefinder $\{S^{(1)}_{3}, S^{(1)}_4\}$, MCG cannot be distinguished from $\Lambda$CDM, and SCG cannot be distinguished from PKK at the present epoch. With $\{\epsilon(z), S^{(1)}_{3}\}$ or $\{\epsilon(z), S_4 \}$, we cannot distinguish PKK from $\Lambda$CDM at the present epoch but the remaining dark energy models are easily distinguished as shown in Figs. \ref{S31e} and \ref{S4e}. SCG cannot be distinguished from PKK at the present epoch by using $\{S_4, S^{(1)}_{4}\}$ but the remaining dark energy models are distinguishable. We can distinguish (but not well) SCG from PKK at the present epoch by using $\{S^{(1)}_{3}, S_{5}\}$ but the remaining dark energy models can be distinguished well from each other.

The results obtained here show that the Statefinder hierarchy and the growth rate of matter perturbations are useful methods with which we can distinguish some dynamical dark energy models from $\Lambda$CDM or even from each other.

\begin{acknowledgments}
This study is supported in part by National Natural Science Foundation of China (Grant Nos. 11147028 and 11273010) and Hebei Provincial Natural Science Foundation of China (Grant No. A2014201068).
\end{acknowledgments}

\end{document}